\documentclass[11pt,twoside]{article}
\usepackage{./asp2010}
\usepackage{graphicx}

\resetcounters

\begin{document}

\title{CoRoT observations of O stars: diverse origins of variability}
\author{R.~Blomme$^1$, M.~Briquet$^2$, P.~Degroote$^2$, L.~Mahy$^3$, 
C.~Aerts$^{2,4}$, J.~Cuypers$^1$, M.~Godart$^3$, E.~Gosset$^3$, 
M.~Hareter$^5$, J.~Montalban$^3$, T.~Morel$^3$, M.F.~Nieva$^{6,7}$, 
A.~Noels$^3$, R.~Oreiro$^2$, E.~Poretti$^8$, N.~Przybilla$^7$, 
M.~Rainer$^8$, G.~Rauw$^3$, F.~Schiller$^7$, S.~Simon-Diaz$^{9,10}$, 
K.~Smolders$^2$, P.~Ventura$^{11}$, M.~Vu\v{c}kovi\'{c}$^2$, 
M.~Auvergne$^{12}$, A.~Baglin$^{12}$, F.~Baudin$^{13}$, C.~Catala$^{12}$, 
E.~Michel$^{12}$ and R.~Samadi$^{12}$
\affil{$^1$Royal Observatory of Belgium, Ringlaan 3, 1180 Brussels, Belgium}
\affil{$^2$Instituut voor Sterrenkunde, K.U. Leuven, Celestijnenlaan 200D,
           3001~Leuven, Belgium}
\affil{$^3$Institut d'Astrophysique et de G\'eophysique, University of
           Li\`ege, B\^at. B5C, All\'ee du 6 Ao\^ut 17, 4000 Li\`ege,
           Belgium}
\affil{$^4$Department of Astrophysics, IMAPP, University of Nijmegen,
           PO Box 9010, 6500 GL Nijmegen, The Netherlands}
\affil{$^5$Institut f\"{u}r Astronomie, Universit\"{a}t Wien,
           T\"{u}rkenschanzstrasse 17, 1180~Vienna, Austria}
\affil{$^6$Max Planck Institute for Astrophysics, Karl Schwarzschild Str. 1, 
           Garching bei M\"{u}nchen 85741, Germany}
\affil{$^7$Dr. Karl Remeis Observatory \& ECAP, University of 
           Erlangen-Nuremberg, Sternwartstrasse 7, 96049 Bamberg, Germany}
\affil{$^8$INAF -- Osservatorio Astronomico di Brera, via E. Bianchi 46, 
           23807 Merate (LC), Italy}
\affil{$^9$Instituto de Astrof\'{\i}sica de Canarias, 38200 La Laguna,
           Tenerife, Spain}
\affil{$^{10}$Departamento de Astrof\'{\i}sica, Universidad de La Laguna, 
             38205 La Laguna, Tenerife, Spain }
\affil{$^{11}$INAF, Osservatorio Astronomico di Roma, Via Frascati 33, 
             00040~Monteporzio Catone (Roma), Italy}
\affil{$^{12}$LESIA, UMR 8109, Observatoire de Paris,
              5 place Jules Janssen, 92195~Meudon Cedex, France}
\affil{$^{13}$Institut d'Astrophysique Spatiale (IAS), B\^atiment 121,
              Universit\'e Paris-Sud, 91405
              Orsay Cedex, France}}

\begin{abstract}
Six O-type stars were observed continuously by the CoRoT satellite during a
34.3-day run. The unprecedented quality of the data allows us to detect even
low-amplitude stellar pulsations in some of these stars (HD~46202 and the
binaries HD~46149 and Plaskett's star). These cover both opacity-driven modes
and solar-like stochastic oscillations, both of importance to the
asteroseismological modelling of O stars. Additional effects can be seen
in the CoRoT light curves, such as binarity and rotational modulation. Some
of the
hottest O-type stars (HD~46223, HD~46150 and HD~46966) are dominated by the
presence of red-noise: we speculate that this is related to a sub-surface
convection zone.
\end{abstract}

\section{Introduction}

The CoRoT satellite \citep{Baglin_2006}
observed the
NGC 2244 cluster and the Mon OB2 association
from 08 Oct to 12 Nov 2008.
As part of this 34.3-day run,
high-precision photometric data were collected every 32 sec on six
O-type stars (see Table~\ref{RB table stars observed}). 
Below we discuss
the light curve analysis of 
each of the stars in turn, going from the latest spectral type
to the earliest one.

\begin{table}[!ht]
\caption{O-type stars observed by CoRoT during run SRa02}
\label{RB table stars observed}
\smallskip
\begin{center}
{\small
\begin{tabular}{lll}
\tableline
\noalign{\smallskip}
star & spectral type & binary period \\
\noalign{\smallskip}
\tableline
\noalign{\smallskip}
HD 46202 & O9 V \\
HD 46149 & O8.5 V + OB-type & $P \approx 2.27$ yr \\
HD 46966 & O8 V \\
HD 47129 & O7.5 I + O6 I & $P=14.39625$ d \\
HD 46150 & O5.5 V ((f)) \\
HD 46223 & O4 V ((f$^{\rm +}$)) \\
\noalign{\smallskip}
\tableline
\end{tabular}
}
\end{center}
\end{table}

\section{Analysis}

\subsection{HD 46202 (O9 V)}

\begin{figure}
\begin{center}
\includegraphics[scale=0.95]{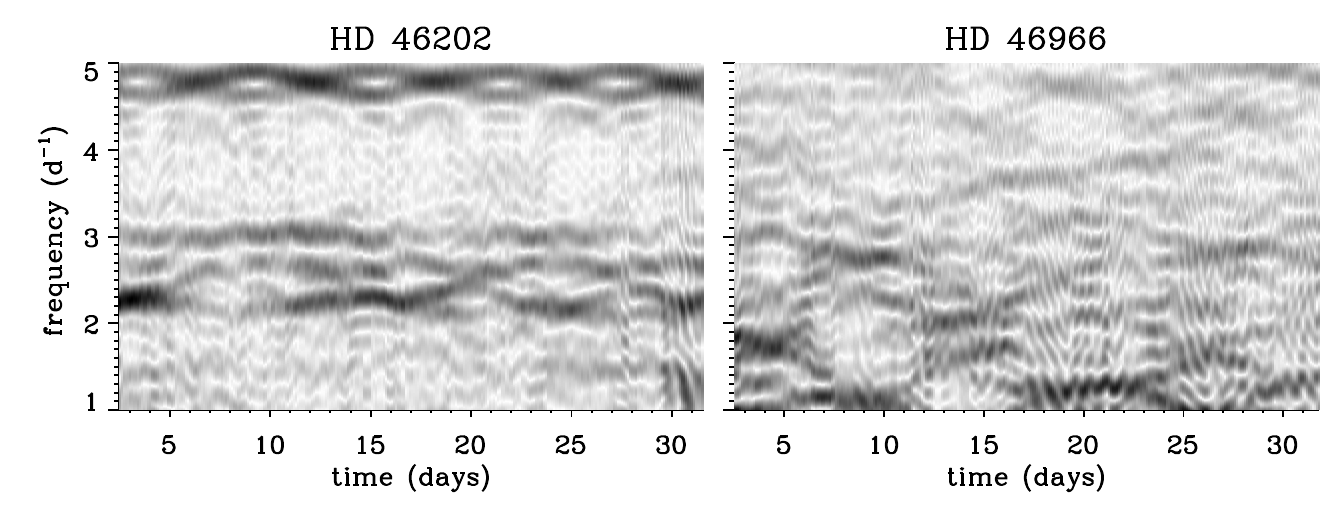}
\end{center}
\caption{Time-frequency diagram of HD~46202 (left) and HD~46966 (right).
The grey scale is proportional to the semi-amplitude, which was calculated
from a Lomb-Scargle analysis on a sliding
5-day window of the observations (centred
on the time given on the x-axis). The time is relative to the start of the
observations.}
\label{RB fig time-frequency}
\end{figure}

This star was studied by \citet{Briquet_2011}. In the periodogram
they found a number of
$\beta$-Cep like pulsation frequencies. The time-frequency diagram
(see Fig.~\ref{RB fig time-frequency}, left) shows that these frequencies
are stable, at least during our campaign,
confirming that they are indeed caused by pulsations
driven by the opacity mechanism.

Asteroseismic modelling of HD~46202 was done 
using the Code Li\'egeois d'\'{E}vo\-lu\-tion Stellaire 
\citep[CL\'ES,][]{Scuflaire_2008}.
The best agreement with the observed frequencies was found for
models in the 23.3 -- 24.9 M$_{\sun}$ range, with a core overshooting 
of 0.05 -- 0.15 times the pressure scale height. The modelling
showed that the
observed modes are not excited in these models, thereby presenting a 
considerable challenge to the theoretical interpretation of the observations.

\subsection{HD 46149 (O8.5 V + OB-type)}

\citet{Degroote_2010} found a rotation period of about 11.8 days in the
light curve of HD~46149, which they attribute to the primary. 
They also found a number of frequencies with a
constant spacing of 0.48 $\pm$ 0.02 d$^{-1}$. This is similar to 
stochastically excited 
p-modes. The time-frequency diagram indicates that these frequencies
are not present during the whole duration of the observing run. A more
detailed analysis shows that they have a lifetime of 3-4 days.

\subsection{HD 46966 (O8 V)}

\citet{Blomme_2011} applied classical prewhitening to the periodogram
of HD~46966. About 300 frequencies are required before the noise level
is reached. Significance tests show that all 300 frequencies are significant.
It is, however, highly suspicious that so many pulsation frequencies
would be present in a single star.
The time-frequency diagram (Fig.~\ref{RB fig time-frequency}, right)
clearly shows that none of these frequencies are stable. They are
therefore not pulsational frequencies. A search for harmonics,
linear combinations or frequency spacings did not turn up any significant
results. Modelling with the ATON evolution code and the MAD pulsation code
predicts pulsations in the 2 -- 4 d$^{-1}$ range, but no clear equivalent
is seen in the observations.

\begin{figure}
\begin{center}
\includegraphics[scale=0.95]{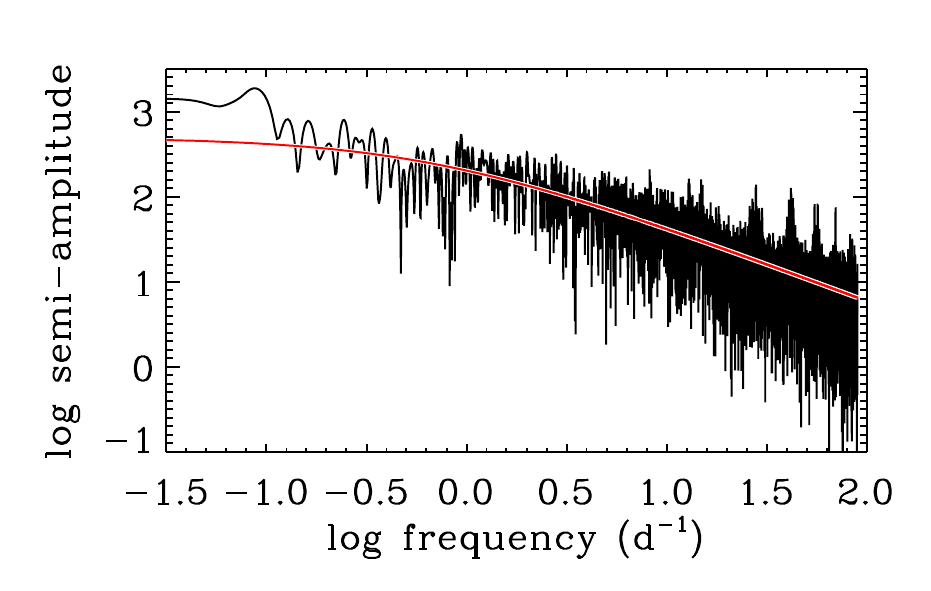}
\end{center}
\caption{Fit (red line) to the red-noise component in the periodogram of 
HD~46966 (black line).}
\label{RB fig red noise}
\end{figure}

All the above points to variations of a more chaotic or stochastic nature.
A good description of this is given by a red-noise spectrum:
red noise power
increases as a power law toward lower frequencies.
In Fig.~\ref{RB fig red noise}, we fit the HD~46966
periodogram with a function that differs slightly from a power law
at lower frequencies.
It is important to stress that this red noise is not an instrumental effect,
because its behaviour is different from
star to star in the simultaneous CoRoT observations discussed here.
Instead, it is caused by the star itself and it indicates stochastic, 
chaotic or 
quasiperiodic effects.
It is seen in a number of other astrophysical
situations, such as the optical
light curves of Mira variables and red supergiants as well as in the 
X-ray light curves of active galaxies, dwarf novae and high-mass X-ray binaries
\citep[see references in][]{Blomme_2011}.

At the moment we can only speculate about the possible causes 
of this red noise in O-type stars. One possibility is that it is caused by
the sub-surface convection zone found in theoretical modelling
\citep{Cantiello_2009}. This zone is assumed to be responsible for a number
of surface effects, such as microturbulence. Other possibilities include
some type of granulation, or the onset of 
clumping in the wind.

\subsection{HD 47129 (O7.5 I + O6 I)}

This is the well-known binary Plaskett's star. \cite{Mahy_2011} found the
orbital period ($P=14.39625$ d) in the observed light curve. They also found 
another significant frequency (0.823 d$^{-1}$) with many harmonics. They 
tentatively ascribe
this frequency to non-radial pulsations (NRPs), possibly generated by the
tidal interaction from both components. The primary and secondary are 
known not to be tidally locked, because they have
very different rotational velocities.
Modelling with the ATON code shows that $l=2,3,4$ NRPs with 
0.8 d$^{-1}$ are indeed possible.

\subsection{HD 46150 (O5.5 V ((f))) and HD 46223 (O4 V ((f$^{\rm +}$)))}

These two stars behave in very similar ways, and are therefore discussed
together. \citet{Blomme_2011} found 500 significant frequencies in their
analysis, but a time-frequency diagram shows that none of these are
stable during the observing run. The ATON evolution code and 
MAD pulsation code were used to predict theoretical frequencies, but
none of these were detected in the observations. Similarly to
HD~46966, a description of the periodogram in terms of red noise
is more appropriate.

\section{Hertzsprung-Russell diagram}

\begin{figure}
\begin{center}
\includegraphics[scale=0.70]{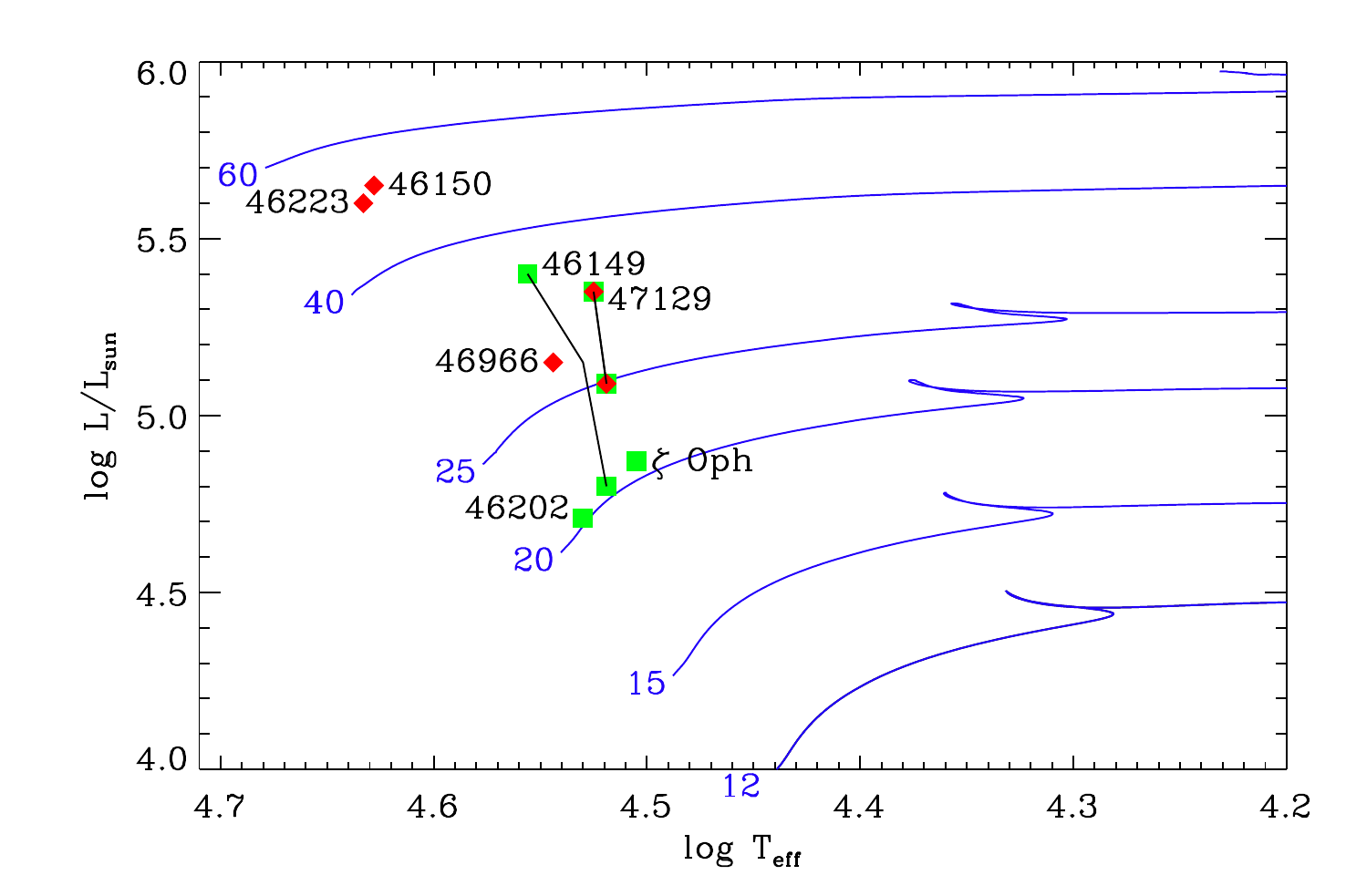}
\end{center}
\caption{HR diagram indicating the position of the stars discussed here,
as well as $\zeta$~Oph. $T_{\rm eff}$ and log $L$ are from the papers cited
in the text. Green squares indicate pulsations, red diamonds
indicate the presence of red noise. Solid lines connect the binary components
for HD~46149 and HD~47129.
The stellar tracks are from \cite{Brott_2011}, with a Zero-Age
Main Sequence rotational
velocity of 220 km\,s$^{-1}$.}
\label{RB fig HR diagram}
\end{figure}

Fig.~\ref{RB fig HR diagram} shows the position of the stars discussed
here in the Hertzsprung-Russell (HR) diagram. We also added $\zeta$~Oph
which shows $\beta$-Cep pulsations in the
MOST photometry \citep{Walker_2005}. We use green squares to indicate 
those stars that show pulsations, and red diamonds for red noise. 
This figure 
describes the extension of the
$\beta$-Cep strip to the hottest stars. The three stars 
HD~46966, HD~46202 and $\zeta$~Oph are especially constraining 
in defining the extension of the $\beta$-Cep strip because
HD~46202 and $\zeta$~Oph show $\beta$-Cep pulsations, but
HD~46966 does not.

\section{Conclusions}

The CoRoT data of six O-type stars show diverse origins for their variability.
They show $\beta$-Cep pulsations (HD~46202),
solar-like oscillations (HD~46149),
the effect of rotation (HD~46149),
the binary period and tidally induced NRPs (HD~47129), as well as
red noise (HD~46966, 46150, 46223).
Modelling the pulsations is possible for HD~46149 and HD~47129,
but the other stars present problems.
Although only six stars were studied, they can help in mapping out
the extension of the $\beta$-Cep strip to the hottest part of HR diagram.
The red noise we find in a number of these stars is intriguing. We cannot
confidently identify its cause, but we speculate that it may be related
to the
sub-surface convection zone expected in these stars, or could be caused by
granulation, or might indicate the onset of 
clumping in the wind. Quantitative modelling of these phenomena 
is urgently required to see which explanation best fits the data.

\bibliography{RBlomme}

\end{document}